\documentclass[%
 reprint,
 amsmath,amssymb,
 aps,
 prl,
]{revtex4-1}

\usepackage[utf8]{inputenc}
\usepackage{graphicx}
\usepackage{dcolumn}
\usepackage{bm}
\usepackage{tikz-feynman}
\usepackage{braket}
\usepackage[colorlinks=true,linktocpage=true,linkcolor=blue,citecolor=blue,allcolors=blue]{hyperref}

\usepackage{color}
\usepackage{xspace}
\usepackage[capitalize]{cleveref}

\newcommand{\UNIT}[1]{\ensuremath{\,{\rm #1}}\xspace}

\newcommand{\MeV}{\UNIT{MeV}}

\newcommand{\fm}{\UNIT{fm}}

\newcommand{\REM}[1]{}

\definecolor{magenta}{cmyk}{0,1,0,0}

\newcommand*{\defeq}{\mathrel{\vcenter{\baselineskip0.5ex \lineskiplimit0pt
                     \hbox{\scriptsize.}\hbox{\scriptsize.}}}%
                     =}

\begin{document}


\title{Open Quantum Systems with Kadanoff-Baym Equations}

\author{Tim Neidig}
\thanks{Corresponding author}
\email{neidig@itp.uni-frankfurt.de}
\affiliation{Institut f\"ur Theoretische Physik, Johann Wolfgang Goethe-Universit\"at, Max-von-Laue-Strasse 1, 60438 Frankfurt am Main, Germany}%
\affiliation{Helmholtz Research Academy Hessen for FAIR (HFHF), GSI Helmholtz Center, Campus Frankfurt, Max-von-Laue-Stra{\ss}e 12, 60438 Frankfurt am Main, Germany}

\author{Jan Rais}
\affiliation{Institut f\"ur Theoretische Physik, Johann Wolfgang Goethe-Universit\"at, Max-von-Laue-Strasse 1, 60438 Frankfurt am Main, Germany}%
\affiliation{Helmholtz Research Academy Hessen for FAIR (HFHF), GSI Helmholtz Center, Campus Frankfurt, Max-von-Laue-Stra{\ss}e 12, 60438 Frankfurt am Main, Germany}

\author{Marcus Bleicher}
\affiliation{Institut f\"ur Theoretische Physik, Johann Wolfgang Goethe-Universit\"at, Max-von-Laue-Strasse 1, 60438 Frankfurt am Main, Germany}%
\affiliation{Helmholtz Research Academy Hessen for FAIR (HFHF), GSI Helmholtz Center, Campus Frankfurt, Max-von-Laue-Stra{\ss}e 12, 60438 Frankfurt am Main, Germany}

\author{Hendrik van Hees}
\affiliation{Institut f\"ur Theoretische Physik, Johann Wolfgang Goethe-Universit\"at, Max-von-Laue-Strasse 1, 60438 Frankfurt am Main, Germany}%
\affiliation{Helmholtz Research Academy Hessen for FAIR (HFHF), GSI Helmholtz Center, Campus Frankfurt, Max-von-Laue-Stra{\ss}e 12, 60438 Frankfurt am Main, Germany}

\author{Carsten Greiner}
\affiliation{Institut f\"ur Theoretische Physik, Johann Wolfgang Goethe-Universit\"at, Max-von-Laue-Strasse 1, 60438 Frankfurt am Main, Germany}%
\affiliation{Helmholtz Research Academy Hessen for FAIR (HFHF), GSI Helmholtz Center, Campus Frankfurt, Max-von-Laue-Stra{\ss}e 12, 60438 Frankfurt am Main, Germany}


\date{\today}

\begin{abstract}
We study the temporal evolution of quantum mechanical fermionic particles exhibiting one bound state within a one-dimensional attractive square-well potential in a heat bath of bosonic particles. For this open quantum system we formulate the non-equilibrium  Kadanoff-Baym equations for the system particles by taking the interactions to be elastic 2-2 scatterings with the heat-bath particles. The corresponding spatially imhomogeneous integro-differential equations for the one-particle Greens's function are solved numerically. We demonstrate how the system particles equilibrate and thermalize with the heat bath and how the off-diagonal elements of the density matrix, expressed in the one-particle energy eigenbasis, decohere, so that only the diagonal entries, i.e. the occupation numbers, survive. In addition, the time evolution of the (retarded) Green's function also determines the spectral properties of the various one-particle quantum states.


\end{abstract}

\maketitle

The coupling of an open quantum system to the environment has been studied over many decades \cite{Weisskopf:1930au, CALDEIRA1983587}. It is highly relevant in the fields of condensed-matter physics, quantum optics, and high-energy physics, but especially in the field of quantum computing, where the life time of a state is limited due to its interaction with the environment and/or q-bit manipulations.
Many approaches to tackle these problems and provide explanations to the decoherence of pure states, employ \textit{Markovian} quantum master equations. It turns out to be not fully understood, why various realisations of master equations do not fulfill fundamental requirements such as the thermalization to the bath temperature, preservation of local conservation laws, complete positivity and norm conservation. 

To be more specific, Markovian quantum master equations, typically in Lindblad form \cite{Gorini:1975nb, 1976CMaPh..48..119L, DIOSI1993517} of the seminal Caldeira-Leggett framework \cite{CALDEIRA1983587, gardiner00, BRE02}, describe the dynamics and all information of the system $S$ (bath $B$, coupling $SB$) in terms of the reduced density matrix, $\hat{\rho}_S$, following the Hamiltonian

\begin{equation}
    \hat{H} = \hat{H}_S + \hat{H}_B + \hat{H}_{SB}, 
\end{equation}

introduced in  \cite{CALDEIRA1983587}. In the following, natural units are used ($c=k_B=\hbar=1$). The master equation, derived from the quantum Liouville equation, is found to take the form 

\begin{equation}
\begin{split}
\label{Lindblad}
    \frac{\partial \hat{\rho}_S}{\partial t} = \text{i} \left[ \hat{\rho}_S, \hat{H}_S\right] + \hat{L}(\hat{\rho}_S, \hat{L}_\lambda),
\end{split}
\end{equation}

where $\hat{L}$ is the Lindblad superoperator, with Lindblad operators $\hat{L}_\lambda)$, which have to be determined for each specific system, and are physicswise ``guided by intuition" \cite{May:1416853}. The advantage of \cref{Lindblad} is that this equation is constructed to fulfill the properties of norm conservation and positivity, in contrast to the Caldeira-Leggett master equation \cite{weiss2008quantum, PhysRevLett.79.3101}. However, Lindblad equations do not necessarily lead to a proper thermalization of the system \cite{PhysRevLett.80.5702}. 
Let us also mention, that the variables in various models describing open quantum systems are often not connected via canonical transformations  \cite{PhysRevA.49.592, PhysRevA.51.1845, Albrecht:1975mtm, GISIN1982364}. The main aim in these kind of approaches is therefore to formulate a correct Langevin-Schrödinger equation or quantum-Langevin equation \cite{10.1063/1.1678812}. 

Alternatively, one can consider non-equilibrium, inhomogenous \textit{non-Markovian} Green's functions propagated through the Kadanoff-Baym equations \cite{KBBook}, as pioneered by Schwinger \cite{10.1063/1.1703727}.
We consider fermionic particles residing in an extended one-dimensional box and experience an attractive square-well potential in the center, which allows for one particular bound state ($E_0 <0$) and otherwise quasi-free states ($E_n >0 \,\, \rm{for} \,\, n \geq 1$) \cite{Rais:2022}. 
The fermions are coupled to a heat bath of free bosonic particles via elastic scatterings in a standard many-body quantum treatment. This represents an open quantum system where energy exchange between the system and bath degrees of freedom forces the system towards a thermal equilibrium which is dictated by the temperature of the bath $T_{\rm{bath}}$, akin to the spirit of the Caldeira-Leggett framework \cite{CALDEIRA1983587}.
In the following we set up the framework for obtaining the resulting Kadanoff-Baym equations and their numerical solution. In particular we will focus on the equilibration, thermalization and decoherence of the various quantum states.

The non-equilibrium Green's function is defined by \cite{10.1063/1.1703727,KBBook, Keldysh:1964ud, Danielewicz:1982kk}

\begin{equation}
\begin{split}
S(1,1') &= -i \bigl\langle T_{c} \bigl[ \hat{\psi}(r,t) \hat{\psi}(r',t')^{\dagger} \bigr] \bigr\rangle \\
&= \Theta_c(t,t') S^{>}(1,1') + \Theta_c(t', t) S^{<}(1,1'), 
\label{1}
\end{split}
\end{equation}

In this notation $1=(r,t)$, $\Theta_c(t,t')$ is the contour-Heaviside function. We assume a spin-saturated system, such that there are no spin dependent interactions, which could make the Green's function non-diagonal in spin-space. $T_c$ is the contour-time ordering operator along the ``Schwinger-Keldysh" contour,

\begin{equation}
\begin{split}
T_{c} \bigl[ \hat{\psi}(r,t) \hat{\psi}(r',t')^{\dagger} \bigr] \defeq \begin{cases}
\hat{\psi}(r,t) \hat{\psi}(r',t')^{\dagger} &\text{if $t > t'$,}\\
\pm \hat{\psi}(r',t')^{\dagger} \hat{\psi}(r,t) &\text{if $t \le t'$}.
\end{cases} 
\label{2}
\end{split}
\end{equation}

The $\pm$ corresponds to bosons/fermions.


We describe fermionic system particles with an external, static, and spatially extended potential coupled to a bosonic environment (heat bath) by the Hamiltonian
\begin{equation}
\begin{split}
 &\hat{H}(t) = \underbrace{\int dr \, \hat{\psi}(r,t)^{\dagger} \Bigl( \underbrace{-\frac{\Delta}{2m_f} + V(r)}_{\defeq h_{0}} \Bigr) \hat{\psi}(r,t)}_{\defeq \hat{H}_{S}(t)} \\
 &+ \underbrace{\lambda \int dr \, \hat{\psi}(r,t)^{\dagger} \hat{\phi}(r,t)^{\dagger} \hat{\psi}(r,t) \hat{\phi}(r,t).}_{\defeq\hat{H}_{\mathrm{SB}}(t)} 
\end{split}
\end{equation}

The bosonic degrees of freedom obey a standard free Hamiltonian. The external potential
 
\begin{equation}
\begin{split}
V(r) \defeq \begin{cases}
-V_{0} &\text{if $|r| \leq \frac{a}{2}$},\\
\,\,\,\,\, 0 &\text{if $|r| > \frac{a}{2}$,}\\
\,\,\,\,\, \infty &\text{if $|r| > \frac{L}{2},$} 
\end{cases} 
\end{split}
\end{equation}
is constructed to mimic a bound state of a particle of size $a$, e.g., a deuteron, or a quantum dot, and an approximated continuum by a sufficiently large square-well extension of size $L\gg a$, gaining the feature of discrete, normalizable and decomposeable states, as introduced in \cite{Rais:2022}. $\hat{\psi}(r,t)$ represents the system-operator and $\hat{\phi}(r,t)$ the bosonic bath-operator. The model employed here is non-relativistic, therefore anti-particles are not considered. $\hat{H}_{\mathrm{SB}}$ represents the local s-wave interaction. We do not  include explicitly self interactions of the system particle or of the bath-particle with itself. This is justified, because the density of bath-particles is much higher, and therefore scatterings of system particles with each other can be neglected. On the other hand we can also think of having only one fermionic particle in the system. 
We understand the bath-particles as a reservoir of constant temperature, $T_{\mathrm{bath}}$, that always stays in equilibrium. This assumption is justified if the bath is much larger than the system we want to study.

We expand the field operators in a basis of one-particle, Schrödinger eigenfunctions of $h_0$ \cite{Keldysh:1964ud, dahlen2006propagating, dahlen2007solving, Stan_2009}, i.e.

\begin{equation}
\begin{split}
h_{0} \phi_{n}(r) &= E_{n} \phi_{n}(r), \\
\int dr \phi_{m}(r)^{*} \phi_{n}(r) &= \delta_{m,n}, \\
\label{eigenfunction}
\end{split}
\end{equation}

via

\begin{equation}
\begin{split}
\hat{\psi}(r,t) &\defeq \sum_{n=0}^{F} \hat{c}_{n}(t) \phi_{n}(r).
\label{4}
\end{split}
\end{equation}

The eigenfunctions can be found in \cite{Rais:2022}, $\phi_0$ represents the bound state, for $n \geq 1$ we have excited, ``free" states. For the numerical solution we truncate the Hilbert space of the modes at a cutoff mode $F$.
Inserting (\ref{4}) in (\ref{1}) leads to the energy-basis representation of the inhomogenous Green's function as

\begin{equation}
\begin{split}
S^{>}(1,1') &= -i \sum_{n,m=0}^{F} \langle \hat{c}_{n}(t) \hat{c}_{m}(t')^{\dagger} \rangle \phi_{n}(r) \phi^{*}_{m}(r'), \\
S^{<}(1,1') &= i \sum_{n,m=0}^{F} \langle \hat{c}_{m}(t')^{\dagger} \hat{c}_{n}(t) \rangle \phi_{n}(r) \phi^{*}_{m}(r').
\label{5}
\end{split}
\end{equation}

These Green's functions are evolved in time by the Kadanoff-Baym equations \cite{KBBook,Danielewicz:1982kk}

\begin{equation}
\begin{split}
\Bigl( i \frac{\partial}{\partial t} + \frac{\Delta_1}{2m_f} -
  V_{\mathrm{eff}}(1) \Bigr) S^{\gtrless}(1,1') &= I_{\mathrm{coll}_{1}}^{\gtrless}(t,t'), 
\label{6.1}
\end{split}
\end{equation}

\begin{equation}
\begin{split}
\Bigl( -i \frac{\partial}{\partial t'} + \frac{\Delta_{1'}}{2m_f} - V_{\mathrm{eff}}(1') \Bigr) S^{\gtrless}(1,1') &= I_{\mathrm{coll}_{2}}^{\gtrless}(t,t'),
\label{6.2}
\end{split}
\end{equation}

where we introduced an effective potential as the sum of the external potential and the Hartree self-energy ($\sim \lambda$)
\begin{equation}
\begin{split}
V_{\mathrm{eff}}(1) &= V(1) + \Sigma_{H}(1), 
\label{effpot}
\end{split}
\end{equation}
and the collision terms on the right-hand side, in which the (second order) direct Born self energy is included,
\begin{equation}
\begin{split}
I_{\mathrm{coll}_{1}}^{\gtrless}(t,t') &= \int_{t_0}^{t} d\bar{1}\biggl[ \Sigma^{>}(1, \bar{1}) - \Sigma^{<}(1, \bar{1}) \biggr] S^{\gtrless}(\bar{1},1') \\
&- \int_{t_0}^{t'} d\bar{1} \Sigma^{\gtrless}(1, \bar{1}) \biggl[S^{>}(\bar{1},1') - S^{<}(\bar{1},1') \biggr], \\
I_{\mathrm{coll}_{2}}^{\gtrless}(t,t') &= \int_{t_0}^{t} d\bar{1} \biggl[S^{>}(1,\bar{1}) - S^{<}(1,\bar{1}) \biggr] \Sigma^{\gtrless}(\bar{1},1’) \\
&- \int_{t_0}^{t’} d\bar{1} S^{\gtrless}(1,\bar{1}) \biggl[ \Sigma^{>}(\bar{1},1’) - \Sigma^{<}(\bar{1},1’) \biggr].
\label{7}
\end{split}
\end{equation}

Before we can define the self-energies explicitly, we first have to introduce the thermal (free)-equilibrium Green's function of the bosonic bath particle
\begin{equation}
\begin{split}
\label{8}
D^{>}_0(1, 1') &= -i \sum_{n}^{B} e^{-i \epsilon_{n}(t-t')} (1+n_{B}(\epsilon_{n})) \tilde{\phi}_{n}(r) \tilde{\phi}^{*}_{n}(r'), \\
D^{<}_0(1, 1') &= -i \sum_{n}^{B} e^{-i \epsilon_{n}(t-t')} n_{B}(\epsilon_{n}) \tilde{\phi}_{n}(r) \tilde{\phi}^{*}_{n}(r'),
\end{split}
\end{equation}
where $\epsilon_{n}=\frac{k_{n}^{2}}{2m_{b}} -\mu_{\mathrm{bath}}$, $k_{n}=\frac{\pi n}{L_{\mathrm{bath}}}$, $n_{B}(\epsilon_{n})=\frac{1}{\mathrm{exp}(\epsilon_{n}/T)-1}$ and $ \tilde{\phi}_{n}(r) = \sqrt{\frac{2}{L_{\mathrm{bath}}}} \, \mathrm{sin}(k_{n}r)$. The chemical potential of the bath fixes the number of particles and $B$ is a cutoff mode similar to the fermionic case.
A dissipative spectral function may also be straightforwardly incorporated in ($\ref{8}$). This is, however, not relevant for our later discussions.

The self-energies being guided by an expansion in the coupling $\lambda$ are given by

\begin{equation}
\begin{split}
\Sigma_{H}(1) &= ( i \lambda) D^{<}_0(1,1^{+}), \\
\Sigma^{\gtrless}(1, 1') &= ( i \lambda)^{2} S^{\gtrless}(1,1') D^{\gtrless}_0(1,1')  D^{\lessgtr}_0(1',1),
\label{9}
\end{split}
\end{equation}

which correspond to the non-dissipative (mean field) tadpole and dissipative sunset diagram. The sunset diagram is more important because it leads to an energy exchange in the scatterings between the bath particles and the system particles, where only a tadpole contribution would not lead to thermal equilibrium.


To solve the Kadanoff-Baym equations numerically, the expansions for the Green's functions (\ref{5}) and (\ref{8}) are inserted into (\ref{6.1}, \ref{6.2}). With the help of (\ref{4}) the set of partial integro-differential equations (\ref{6.1}, \ref{6.2}) reduce to a set of ordinary integro-differential equations for the matrix-valued expectation values $c^{>}_{n,m}(t,t’) \defeq \langle \hat{c}_{n}(t) \hat{c}_{m}(t')^{\dagger} \rangle $ and $c^{<}_{n,m}(t,t’) \defeq \langle \hat{c}_{m}(t')^{\dagger} \hat{c}_{n}(t) \rangle $ in the two-time plane. The self-energies can be expanded in the same basis functions
\begin{equation}
\begin{split}
\label{10}
 & \Sigma^{\gtrless}(1, 1') = \mp i \sum_{b,a}^{F} \Sigma^{\gtrless}_{b,a}(t,t’) \phi_{b}(r) \phi^{*}_{a}(r'), \\
 & \Sigma^{\gtrless}_{b,a}(t,t’) = \lambda^{2} \sum_{n,m}^{F} \Bigl( \sum_{j,k}^{B} e^{\mp i (\epsilon_{j}-\epsilon_{k})(t-t’)} \, (1+n_{B}(\epsilon_{j})) \, n_{B}(\epsilon_{k}) \\
 & \underbrace{\int dr \phi^{*}_{b}(r) \phi_{n}(r) \tilde{\phi}_{j}(r)\tilde{\phi}^{*}_{k}(r)}_{\defeq V_{b,n,j,k}} \, c^{\gtrless}_{n,m}(t,t’) \, V_{m,a,k,j}  \Bigr), \\
 &\Sigma_{H_{b,a}} = \lambda \sum_{j}^{B} e^{-i \epsilon_{j}0^{+}} n_{B}(\epsilon_{j}) V_{b,a,j,j}.
\end{split}
\end{equation}
The coefficients $c^{\gtrless}_{n,m}(t,t’)$ are propagated in the discretized two-time plane according to (\ref{6.1}, \ref{6.2}). Not all four equations are needed, because the Green's functions are skew-hermitian,
\begin{equation}
S^{\gtrless}(1,1’) = -S^{\gtrless}(1’,1)^{\dagger} \rightarrow
  c^{\gtrless}_{n,m}(t,t’) = c^{\gtrless *}_{m,n}(t’,t).
\label{11}
\end{equation}
Thus it is only necessary to calculate $c^{\gtrless}_{n,m}(t,t’)$ in the corresponding upper/lower triangle of the two-time plane \cite{DANIELEWICZ1984305, KOHLER1999123, Juchem_2004, dahlen2006propagating, dahlen2007solving, Stan_2009, PhysRevB.82.155108, Schenke:2005ry, Meirinhos_2022}. Consequently, only (\ref{6.1}) is used for $c^{>}_{n,m}(t,t’)$ in $t$-direction and (\ref{6.2}) for $c^{<}_{n,m}(t,t’)$ in $t’$-direction. For the time-diagonal instead only $c^{<}_{n,m}(t,t)$ is propagated through a combination of both Eqs. (\ref{6.1}, \ref{6.2}), and  $c^{>}_{n,m}(t,t)$ is then fixed via the (equal-time) anti-commutation relation for fermions, $c^{>}_{n,m}(t,t) + c^{<}_{n,m}(t,t) = \delta_{n,m} $. The resulting equations are
\begin{equation}
\begin{split}
&\frac{\partial}{\partial t} c^{>}_{n,m}(t,t’) + i \sum_{i}^{F}
  V_{\mathrm{eff}n,i}(t) c^{>}_{i,m}(t,t’) = I^{>}_{n,m 1}(t,t'), \\
&\frac{\partial}{\partial t’} c^{<}_{n,m}(t,t’) - i \sum_{i}^{F}
  c^{<}_{n,i}(t,t’) V_{\mathrm{eff}i,m}(t') = I^{<}_{n,m 2}(t,t'), \\
-&\frac{\partial}{\partial t} c^{<}_{n,m}(t,t) + i [c^{<}, V_{\mathrm{eff}}
   ]_{n,m}(t) = I^{<}_{n,m 1}(t,t) - I^{<}_{n,m 2}(t,t) 
\label{12}
\end{split}
\end{equation}
with 
\begin{equation}
\begin{split}
V_{\mathrm{eff}n,m}(t) &= E_n \delta_{n,m} + \Sigma_{Hn,m},  \\
[c^{<}, V_{\mathrm{eff}} ]_{n,m}(t) &= \sum_{i}^{F} c^{<}_{n,i}(t,t) V_{\mathrm{eff}i,m}(t) - V_{\mathrm{eff}n,i}(t) c^{<}_{i,m}(t,t) , 
\end{split}
\end{equation}
and
\begin{equation}
\begin{split}
I^{>}_{n,m 1}(t,t') &= -\int_{t_0}^{t} d\bar{t} \sum_{i}^{F} \biggl[ \Sigma^{>}_{n,i}(t,\bar{t}) + \Sigma^{<}_{n,i}(t,\bar{t}) \biggr] c^{>}_{i,m}(\bar{t},t')  \\
&+ \int_{t_0}^{t'} d\bar{t} \sum_{i}^{F} \Sigma^{>}_{n,i}(t,\bar{t}) \biggl[c^{>}_{i,m}(\bar{t},t') + c^{<}_{i,m}(\bar{t},t') \biggr] , \\
I^{<}_{n,m 1}(t,t') &= \int_{t_0}^{t} d\bar{t} \sum_{i}^{F} \biggl[ \Sigma^{>}_{n,i}(t,\bar{t}) + \Sigma^{<}_{n,i}(t,\bar{t}) \biggr] c^{<}_{i,m}(\bar{t},t')  \\
&- \int_{t_0}^{t'} d\bar{t} \sum_{i}^{F} \Sigma^{<}_{n,i}(t,\bar{t}) \biggl[c^{>}_{i,m}(\bar{t},t') + c^{<}_{i,m}(\bar{t},t') \biggr] , \\
I^{<}_{n,m 2}(t,t') &= \int_{t_0}^{t} d\bar{t} \sum_{i}^{F} \biggl[c^{>}_{n,i}(t,\bar{t}) + c^{<}_{n,i}(t,\bar{t}) \biggr] \Sigma^{<}_{i,m}(\bar{t},t')  \\
&-\int_{t_0}^{t'} d\bar{t} \sum_{i}^{F} c^{<}_{n,i}(t,\bar{t}) \biggl[ \Sigma^{>}_{i,m}(\bar{t},t') + \Sigma^{<}_{i,m}(\bar{t},t') \biggr] .
\label{13}
\end{split}
\end{equation}

The solution to this highly coupled system of ordinary integro-differential equations, is obtained numerically. We apply a predictor-corrector method known as ``Heun's method", which has already been used in previous works \cite{DANIELEWICZ1984305, KOHLER1999123, Juchem_2004, dahlen2006propagating, dahlen2007solving, Stan_2009, PhysRevB.82.155108, Meirinhos_2022}, (in many cases even the explicit ``Euler method" gives sufficient accuracy). 
For a detailed overview of the numerical treatment we refer the reader to \cite{BalzerPHD, PhysRevA.82.033427}. 

For the following calculations, the model parameters are: $m_{b}=138 \MeV$, $m_{f}=938 \MeV$, $L=20 \fm$, $L_{\mathrm{bath}}=55 \fm$, $T_{\mathrm{bath}}= 100 \MeV$, $\mu_{\mathrm{bath}} = 0 \MeV$, $\lambda=0.35$, $V_{0}=12.8 \MeV $, $a= 1.2 \fm$, $N_{t}=1400$, $\delta t=0.075 \fm$, $F=25$ and $B=30$. We have checked the dependence on the discretization value and the cut-off parameters. For the bound state we have a binding energy of $E_0 \sim -2.23 \MeV$ \cite{Rais:2022}. We neglect the Hartree term in the following calculations, because it only leads to a $T_{\mathrm{bath}}$ and $\mu_{\mathrm{bath}}$ dependent shift of the energy levels, but it does not contribute to the thermalization process due to scatterings. The coefficients $c^{<}_{n,m}(0,0)$, i.e., the initial density matrix, have to be specified to describe the initial fermionic system at time $t=0$. The diagonal elements $c^{<}_{n,n}(t,t)$ define the occupation number of the corresponding state, whereas the off diagonal entries specify quantum correlations among various states.  

\textit{Spectral properties.}
Unlike as in the Master-equation
formalism, the Green's function provides spectral information of the
energy distribution of all particle states of the system during the
non-equilibrium, time evolution and not only for the long-time
equilibrium information. For this purpose we define the spectral coefficients as $a_{n,m}(t,t') \defeq c^{>}_{n,m}(t,t') + c^{<}_{n,m}(t,t')$. In the following, we introduce a central time, $\bar{T} = \frac{t+t'}{2}$, and a relative time, $\Delta{t} = t-t'$ \cite{KBBook, Danielewicz:1982kk, Juchem_2004} (other parameterizations are possible \cite{Schenke:2005ry}). These spectral coefficients oscillate in relative time by a characteristic frequency, which can be interpreted as the quasi-particle energy of this state. To make this clearer, we introduce the Wigner transform of the spectral coefficients as

\begin{equation}
\tilde{a}_{n,m}(\omega,\bar{T}) = \int d\Delta{t} \, e^{i \omega \Delta{t}} a_{n,m}(\bar{T}+\frac{\Delta{t}}{2},\bar{T}-\frac{\Delta{t}}{2}).
\label{14}
\end{equation}

In the no-interaction case, we can directly see from (\ref{8}), that the Wigner transform is just a $\delta$-distribution at frequency $\omega = \epsilon_{n}$. When we switch on the interaction, the finite self-energies lead to a shift of the peak (real part of the retarded self-energy) and a broadening of the spectral function (due to the imaginary part of the retarded self-energy), where the width $\Gamma$ can be understood as an inverse life time of the state \cite{KBBook, Danielewicz:1982kk, Juchem_2004}:

\begin{equation}
\begin{split}
& \Gamma_{n,m}(\bar{T},\omega) = -2 \, \rm{Im}(\Sigma^{\mathrm{ret}}_{n,m}(\bar{T},\omega)) = \int d\Delta{t} \, e^{i \omega \Delta{t}} \\
&\Bigl[ \Bigl( \Sigma^{>}_{n,m}\Bigl(\bar{T}+\frac{\Delta{t}}{2},\bar{T}-\frac{\Delta{t}}{2}\Bigr) 
+ \Sigma^{<}_{n,m}\Bigl(\bar{T}+\frac{\Delta{t}}{2},\bar{T}-\frac{\Delta{t}}{2}\Bigr) \Bigr) \Bigr],
\label{15.1}
\end{split}
\end{equation}

\begin{equation}
\begin{split}
& \rm{Re}(\Sigma^{\mathrm{ret}}_{n,m}(\bar{T},\omega)) = \frac{-i}{2} \int d\Delta{t} \, e^{i \omega \Delta{t}} \Bigl[ sign(\Delta{t}) \\ &\Bigl( \Sigma^{>}_{n,m}\Bigl(\bar{T}+\frac{\Delta{t}}{2},\bar{T}-\frac{\Delta{t}}{2}\Bigr)
+ \Sigma^{<}_{n,m}\Bigl(\bar{T}+\frac{\Delta{t}}{2},\bar{T}-\frac{\Delta{t}}{2}\Bigr) \Bigr) \Bigr].
\label{15.2}
\end{split}
\end{equation}

\begin{figure}[]
  \begin{center}
    \hspace*{\fill}%
    \includegraphics[width=1.15\columnwidth,clip=true]{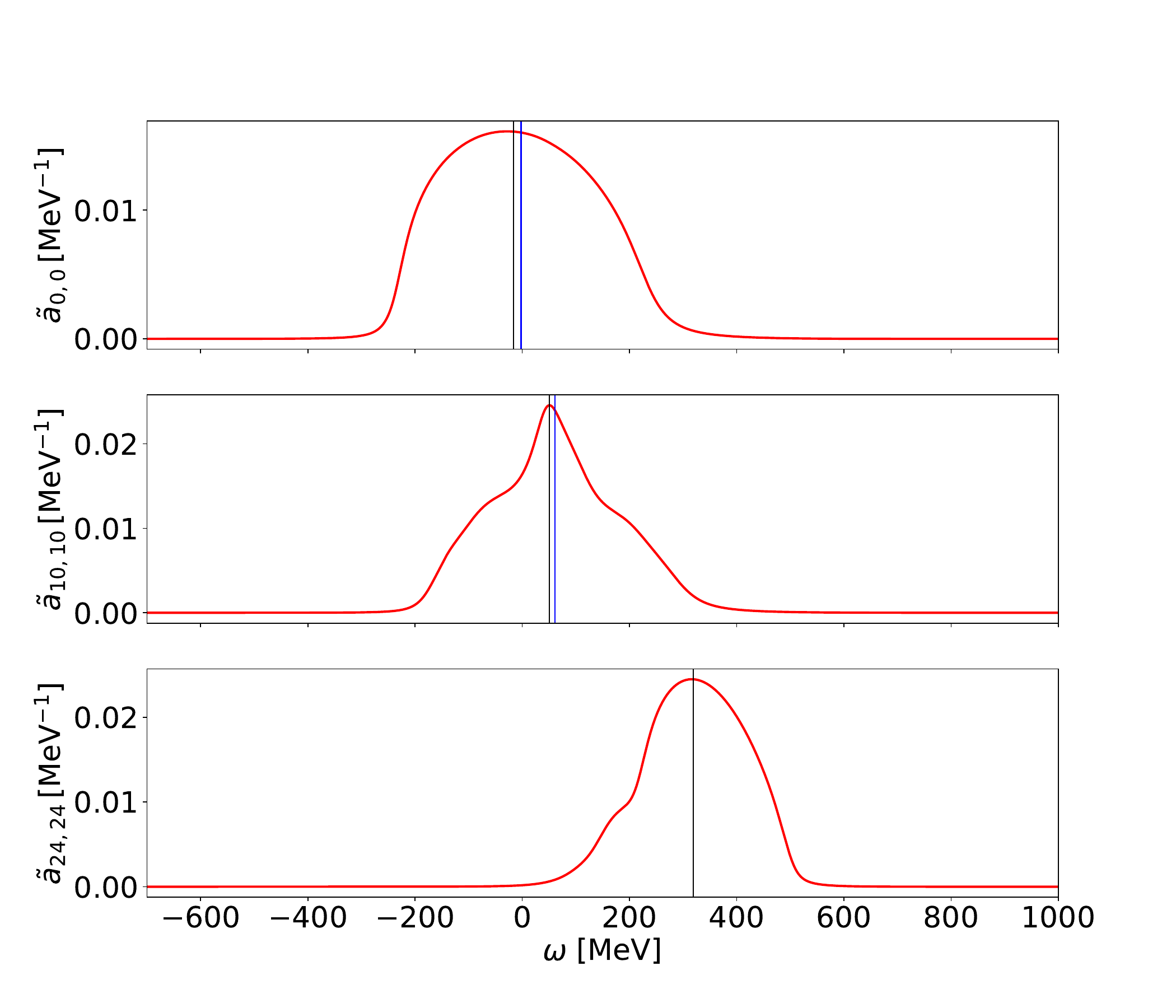}
    \hspace*{\fill}%

    \caption{
    Spectral functions $\tilde{a}_{0,0}(\omega,\bar{T}=52 \mathrm{fm})$, $\tilde{a}_{10,10}(\omega,\bar{T}=52 \mathrm{fm})$ and $\tilde{a}_{24,24}(\omega,\bar{T}=52 \mathrm{fm})$.
    }
    \label{fig:2}
  \end{center}
\end{figure}

\begin{figure}[]
  \begin{center}
    \hspace*{\fill}%
    \includegraphics[width=1.1\columnwidth,clip=true]{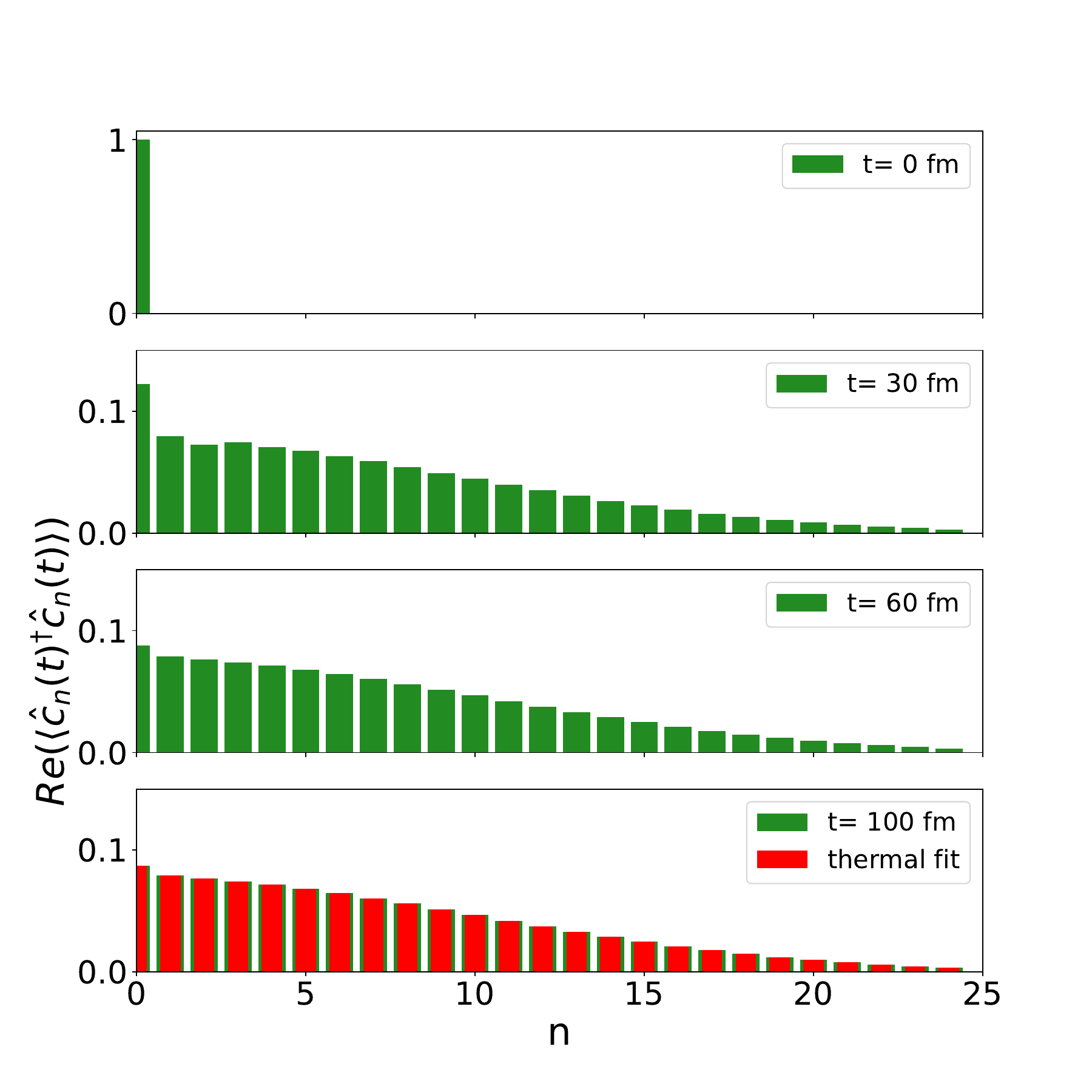}
    \hspace*{\fill}%
    \caption{
      $c^{<}_{n,n}(t,t)$ plotted for different times. The occupation number of the final states ($t=100 \mathrm{fm}$) was fitted to a Fermi-Dirac distribution yield $T_{\mathrm{system}}\approx 100.133 \MeV$ and $\mu_{\mathrm{system}} \approx -298.125 \MeV$.}
    \label{fig:3}
  \end{center}
\end{figure}

In Fig. \ref{fig:2} the spectral functions of three different states are shown, the bound state and two continuum states. For this particular example, we evolve the system for the initial occupation number depicted in Fig. \ref{fig:3}. The off diagonal elements are taken to be zero. For the central time we take $20 \fm$. The spectral width is about $250 \MeV$, and decreases for higher states. The two slightly separate lines in the plots show the position of the ``on-shell" energy $E_{n}$ (blue line) and the ``in-medium" energy $E_{n} + \mathrm{Re}(\Sigma^{\text{ret}}_{n,n}(\bar{T},\omega))$ (black line). Again we observe, that the shift of the peak gets smaller for higher states.

\textit{Equilibration and Thermalization.}
The process of equilibration and thermalization of the system can now be discussed. In the long-time limit the system should approach a thermal equilibrium fixed point defined by $T_{\mathrm{bath}}$. 
In this limit, the diagonal elements $c^{<}_{n,n}(t,t)$ should approach the Fermi-Dirac distribution

\begin{equation}
\begin{split}
\mathrm{lim}_{t \rightarrow \infty} c^{<}_{n,n}(t,t) &= \int \frac{d\mathrm{\omega}}{2\pi} \, n_{\mathrm{F}}(T_{\mathrm{syst}}, \mu_{\mathrm{syst}}, \omega) \, \tilde{a}_{n,n}(\omega,\bar{T} )  
\label{16}
\end{split}
\end{equation}

In Fig. \ref{fig:3}, we show the evolution of the occupation numbers $c^{<}_{n,n}(t,t)$ from the initial conditions through intermediate times until full equilibration has emerged. The temperature $T_{\mathrm{syst}}=T_{\mathrm{bath}}$ is reached independently of the coupling or initial conditions, but the chemical potential in thermal equilibrium depends, of course on the total number of fermions. 
This originates from the fact, that the Kadanoff-Baym equations are conserving particle number (and energy/momentum for closed systems), if the self-energies are $\Phi$-derivable \cite{PhysRev.124.287,PhysRev.127.1391,DANIELEWICZ1990154}. 

\textit{Decoherence}.
All numerical simulations show, that the off-diagonal elements, if initially non-zero, vanish for long times. 
Thus, we want to emphasize, that the feature of quantum decoherence is automatically included in the Kadanoff-Baym equations. To demonstrate this more concretely, we  consider one single fermion being initially in a pure, superimposed state of two levels, i.e. $\ket{\phi}_{\text{super}}=\frac{1}{\sqrt{2}}(\ket{\phi_{10}} + \ket{\phi_{15}})$, and alternatively a single level in the ground state $\ket{\phi}_{\text{ground}}=\ket{\phi_{0}} $.
For the superimposed state, one has $c^{<}_{10,10}(0,0) = c^{<}_{15,10}(0,0) = c^{<}_{10,15}(0,0) = c^{<}_{15,15}(0,0) = 0.5$, and for the single ground state we have $c^{<}_{0,0}(0,0) = 1.0$. 

\begin{figure}[]
    \hspace*{\fill}%
    \includegraphics[width=1.1\columnwidth,clip=true]{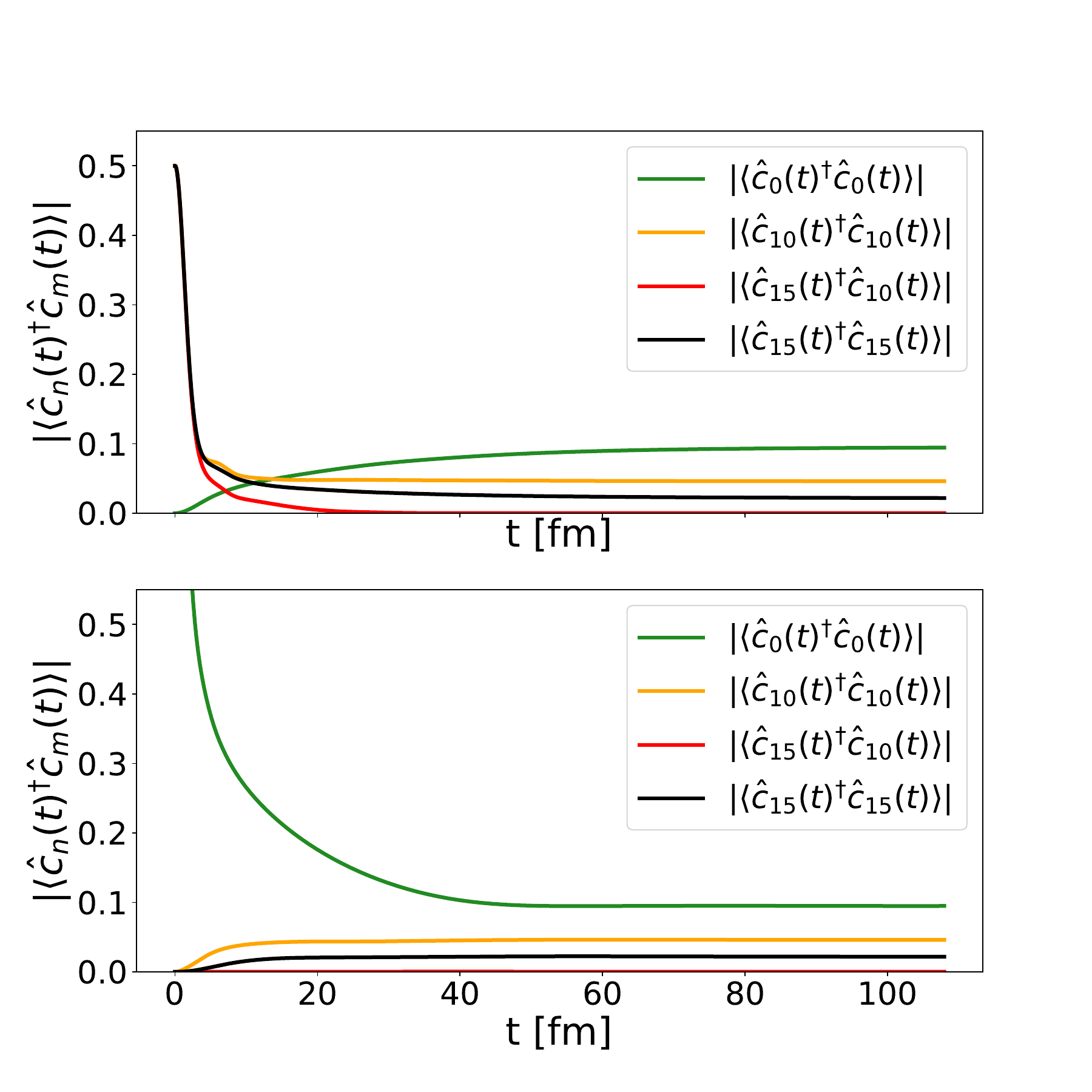}
    \hspace*{\fill}%

    \caption{
     The time evolution of an initial two-level system (top) and an initial single level (bottom) plotted for the relevant matrix elements.
    }
    \label{fig:4}
\end{figure}

In Fig. \ref{fig:4}, the decay of both pure states into a mixed state is shown. The off-diagonal element $c^{<}_{10,15}(t,t) \defeq \langle \hat{c}_{15}(t)^{\dagger} \hat{c}_{10}(t) \rangle $ for the two-level system dies out quickly due to the ongoing dephasing of random scatterings of the fermion with the bath particles.
One can see, that the off-diagonal elements first decay before the system finally equilibrates (FIG. \ref{fig:4}, top). The equilibration time towards thermal equilibrium for all occupation numbers is larger than in the case of initially vanishing off-diagonal elements on a timescale dictated by the coupling, temperature and density. In principle, the phenomenon of decoherence is the foundation to allow the use of more simplistic kinetic-quantum-Boltzmann equations or standard Master-equations.

\textit{Summary and Conclusions.}
Green's functions have been studied for homogeneous systems in \cite{DANIELEWICZ1984305, KOHLER1999123, Juchem_2004} and also for inhomogeneous systems in \cite{dahlen2006propagating, dahlen2007solving, Stan_2009, PhysRevA.82.033427, PhysRevB.82.155108, PhysRevD.88.085009}.
The question of equilibration in a closed quantum system was discussed in the homogeneous case in \cite{Juchem_2004}, but in an open quantum system for the inhomogeneous case equilibration and decoherence have not been analyzed in this framework. 
We have shown for the first time that the
framework of the Kadanoff-Baym equations is a promising and powerful alternative to the widely used
Lindblad formalism for solving
nonequilibrium problems in open quantum systems.
The Kadanoff-Baym equations preserve local conservation laws, the norm and complete positivity of the occupation numbers by construction, if the self-energies are $\Phi$-derivable as implemented in our approach \cite{PhysRev.124.287,PhysRev.127.1391,DANIELEWICZ1990154}. In an open quantum system, thermalization is reached for the dressed quantum degrees of freedom and quantum correlations decohere by the same interactions among the system and bath degrees of freedom. We have demonstrated these features for fermionic particles (e.g., nucleons) with an external potential to produce a bound state ($E_0 < 0$) in the presence of lighter bosonic particles (e.g. pions) in one spatial dimension. The extension to three dimensions is straightforward, in order to demonstrate how bound states are formed, populated, depopulated, and how they thermalize in an open quantum system \cite{workinprogress}. Such an understanding is e.g.  relevant for the formation of deuterons \cite{Danielewicz:1991dh,Vovchenko:2019aoz,Neidig:2021bal,Rais:2022,Coci:2023daq} or heavy quarkonia ($J/\Psi$'s) \cite{Beraudo:2010tw} in a relativistic heavy ion collision. 


The authors acknowledge for the support the European Union's Horizon 
2020 research and innovation program
under grant agreement No 824093 (STRONG-2020).
C.G. thanks Volodymyr Vovchenko for fruitful discussions.
T.N. is supported by Helmholtz Forschungsakademie Hessen für FAIR (HFHF).
We acknowledge support by the Deutsche Forschungsgemeinschaft (DFG) through the CRC-TR 211 ``Strong-interaction matter under extreme conditions". 
All calculations have been done on the Goethe-HLR (CSC) in Frankfurt am Main.

\appendix



\bibliography{references}%

\begin{thebibliography}{42}%
\makeatletter
\providecommand \@ifxundefined [1]{%
 \@ifx{#1\undefined}
}%
\providecommand \@ifnum [1]{%
 \ifnum #1\expandafter \@firstoftwo
 \else \expandafter \@secondoftwo
 \fi
}%
\providecommand \@ifx [1]{%
 \ifx #1\expandafter \@firstoftwo
 \else \expandafter \@secondoftwo
 \fi
}%
\providecommand \natexlab [1]{#1}%
\providecommand \enquote  [1]{``#1''}%
\providecommand \bibnamefont  [1]{#1}%
\providecommand \bibfnamefont [1]{#1}%
\providecommand \citenamefont [1]{#1}%
\providecommand \href@noop [0]{\@secondoftwo}%
\providecommand \href [0]{\begingroup \@sanitize@url \@href}%
\providecommand \@href[1]{\@@startlink{#1}\@@href}%
\providecommand \@@href[1]{\endgroup#1\@@endlink}%
\providecommand \@sanitize@url [0]{\catcode `\\12\catcode `\$12\catcode
  `\&12\catcode `\#12\catcode `\^12\catcode `\_12\catcode `\%12\relax}%
\providecommand \@@startlink[1]{}%
\providecommand \@@endlink[0]{}%
\providecommand \url  [0]{\begingroup\@sanitize@url \@url }%
\providecommand \@url [1]{\endgroup\@href {#1}{\urlprefix }}%
\providecommand \urlprefix  [0]{URL }%
\providecommand \Eprint [0]{\href }%
\providecommand \doibase [0]{http://dx.doi.org/}%
\providecommand \selectlanguage [0]{\@gobble}%
\providecommand \bibinfo  [0]{\@secondoftwo}%
\providecommand \bibfield  [0]{\@secondoftwo}%
\providecommand \translation [1]{[#1]}%
\providecommand \BibitemOpen [0]{}%
\providecommand \bibitemStop [0]{}%
\providecommand \bibitemNoStop [0]{.\EOS\space}%
\providecommand \EOS [0]{\spacefactor3000\relax}%
\providecommand \BibitemShut  [1]{\csname bibitem#1\endcsname}%
\let\auto@bib@innerbib\@empty
\bibitem [{\citenamefont {Weisskopf}\ and\ \citenamefont
  {Wigner}(1930)}]{Weisskopf:1930au}%
  \BibitemOpen
  \bibfield  {author} {\bibinfo {author} {\bibfnamefont {V.}~\bibnamefont
  {Weisskopf}}\ and\ \bibinfo {author} {\bibfnamefont {E.~P.}\ \bibnamefont
  {Wigner}},\ }\href {\doibase 10.1007/BF01336768} {\bibfield  {journal}
  {\bibinfo  {journal} {Z. Phys.}\ }\textbf {\bibinfo {volume} {63}},\ \bibinfo
  {pages} {54} (\bibinfo {year} {1930})}\BibitemShut {NoStop}%
\bibitem [{\citenamefont {Caldeira}\ and\ \citenamefont
  {Leggett}(1983)}]{CALDEIRA1983587}%
  \BibitemOpen
  \bibfield  {author} {\bibinfo {author} {\bibfnamefont {A.}~\bibnamefont
  {Caldeira}}\ and\ \bibinfo {author} {\bibfnamefont {A.}~\bibnamefont
  {Leggett}},\ }\href {\doibase https://doi.org/10.1016/0378-4371(83)90013-4}
  {\bibfield  {journal} {\bibinfo  {journal} {Physica A: Statistical Mechanics
  and its Applications}\ }\textbf {\bibinfo {volume} {121}},\ \bibinfo {pages}
  {587} (\bibinfo {year} {1983})}\BibitemShut {NoStop}%
\bibitem [{\citenamefont {Gorini}\ \emph {et~al.}(1976)\citenamefont {Gorini},
  \citenamefont {Kossakowski},\ and\ \citenamefont
  {Sudarshan}}]{Gorini:1975nb}%
  \BibitemOpen
  \bibfield  {author} {\bibinfo {author} {\bibfnamefont {V.}~\bibnamefont
  {Gorini}}, \bibinfo {author} {\bibfnamefont {A.}~\bibnamefont {Kossakowski}},
  \ and\ \bibinfo {author} {\bibfnamefont {E.~C.~G.}\ \bibnamefont
  {Sudarshan}},\ }\href {\doibase 10.1063/1.522979} {\bibfield  {journal}
  {\bibinfo  {journal} {J. Math. Phys.}\ }\textbf {\bibinfo {volume} {17}},\
  \bibinfo {pages} {821} (\bibinfo {year} {1976})}\BibitemShut {NoStop}%
\bibitem [{\citenamefont {{Lindblad}}(1976)}]{1976CMaPh..48..119L}%
  \BibitemOpen
  \bibfield  {author} {\bibinfo {author} {\bibfnamefont {G.}~\bibnamefont
  {{Lindblad}}},\ }\href {\doibase 10.1007/BF01608499} {\bibfield  {journal}
  {\bibinfo  {journal} {Communications in Mathematical Physics}\ }\textbf
  {\bibinfo {volume} {48}},\ \bibinfo {pages} {119} (\bibinfo {year}
  {1976})}\BibitemShut {NoStop}%
\bibitem [{\citenamefont {Diósi}(1993)}]{DIOSI1993517}%
  \BibitemOpen
  \bibfield  {author} {\bibinfo {author} {\bibfnamefont {L.}~\bibnamefont
  {Diósi}},\ }\href {\doibase https://doi.org/10.1016/0378-4371(93)90065-C}
  {\bibfield  {journal} {\bibinfo  {journal} {Physica A}\ }\textbf {\bibinfo
  {volume} {199}},\ \bibinfo {pages} {517} (\bibinfo {year}
  {1993})}\BibitemShut {NoStop}%
\bibitem [{\citenamefont {Gardiner}\ and\ \citenamefont
  {Zoller}(2000)}]{gardiner00}%
  \BibitemOpen
  \bibfield  {author} {\bibinfo {author} {\bibfnamefont {C.~W.}\ \bibnamefont
  {Gardiner}}\ and\ \bibinfo {author} {\bibfnamefont {P.}~\bibnamefont
  {Zoller}},\ }\href@noop {} {\emph {\bibinfo {title} {Quantum Noise}}},\
  \bibinfo {edition} {2nd}\ ed.,\ edited by\ \bibinfo {editor} {\bibfnamefont
  {H.}~\bibnamefont {Haken}}\ (\bibinfo  {publisher} {Springer},\ \bibinfo
  {year} {2000})\BibitemShut {NoStop}%
\bibitem [{\citenamefont {Breuer}\ and\ \citenamefont
  {Petruccione}(2002)}]{BRE02}%
  \BibitemOpen
  \bibfield  {author} {\bibinfo {author} {\bibfnamefont {H.~P.}\ \bibnamefont
  {Breuer}}\ and\ \bibinfo {author} {\bibfnamefont {F.}~\bibnamefont
  {Petruccione}},\ }\href@noop {} {\emph {\bibinfo {title} {The theory of open
  quantum systems}}}\ (\bibinfo  {publisher} {Oxford University Press},\
  \bibinfo {address} {Great Clarendon Street},\ \bibinfo {year}
  {2002})\BibitemShut {NoStop}%
\bibitem [{\citenamefont {May}\ and\ \citenamefont
  {Kühn}(2011)}]{May:1416853}%
  \BibitemOpen
  \bibfield  {author} {\bibinfo {author} {\bibfnamefont {V.}~\bibnamefont
  {May}}\ and\ \bibinfo {author} {\bibfnamefont {O.}~\bibnamefont {Kühn}},\
  }\href {http://cds.cern.ch/record/1416853} {\emph {\bibinfo {title} {{Charge
  and Energy Transfer Dynamics in Molecular Systems}}}}\ (\bibinfo  {publisher}
  {John Wiley \& Sons},\ \bibinfo {address} {Hoboken, NJ},\ \bibinfo {year}
  {2011})\BibitemShut {NoStop}%
\bibitem [{\citenamefont {Weiss}(2008)}]{weiss2008quantum}%
  \BibitemOpen
  \bibfield  {author} {\bibinfo {author} {\bibfnamefont {U.}~\bibnamefont
  {Weiss}},\ }\href {https://books.google.de/books?id=mGVhDQAAQBAJ} {\emph
  {\bibinfo {title} {Quantum Dissipative Systems}}},\ Series in modern
  condensed matter physics\ (\bibinfo  {publisher} {World Scientific},\
  \bibinfo {year} {2008})\BibitemShut {NoStop}%
\bibitem [{\citenamefont {Gao}(1997)}]{PhysRevLett.79.3101}%
  \BibitemOpen
  \bibfield  {author} {\bibinfo {author} {\bibfnamefont {S.}~\bibnamefont
  {Gao}},\ }\href {\doibase 10.1103/PhysRevLett.79.3101} {\bibfield  {journal}
  {\bibinfo  {journal} {Phys. Rev. Lett.}\ }\textbf {\bibinfo {volume} {79}},\
  \bibinfo {pages} {3101} (\bibinfo {year} {1997})}\BibitemShut {NoStop}%
\bibitem [{\citenamefont {Wiseman}\ and\ \citenamefont
  {Munro}(1998)}]{PhysRevLett.80.5702}%
  \BibitemOpen
  \bibfield  {author} {\bibinfo {author} {\bibfnamefont {H.~M.}\ \bibnamefont
  {Wiseman}}\ and\ \bibinfo {author} {\bibfnamefont {W.~J.}\ \bibnamefont
  {Munro}},\ }\href {\doibase 10.1103/PhysRevLett.80.5702} {\bibfield
  {journal} {\bibinfo  {journal} {Phys. Rev. Lett.}\ }\textbf {\bibinfo
  {volume} {80}},\ \bibinfo {pages} {5702} (\bibinfo {year}
  {1998})}\BibitemShut {NoStop}%
\bibitem [{\citenamefont {Yu}\ and\ \citenamefont
  {Sun}(1994)}]{PhysRevA.49.592}%
  \BibitemOpen
  \bibfield  {author} {\bibinfo {author} {\bibfnamefont {L.~H.}\ \bibnamefont
  {Yu}}\ and\ \bibinfo {author} {\bibfnamefont {C.-P.}\ \bibnamefont {Sun}},\
  }\href {\doibase 10.1103/PhysRevA.49.592} {\bibfield  {journal} {\bibinfo
  {journal} {Phys. Rev. A}\ }\textbf {\bibinfo {volume} {49}},\ \bibinfo
  {pages} {592} (\bibinfo {year} {1994})}\BibitemShut {NoStop}%
\bibitem [{\citenamefont {Sun}\ and\ \citenamefont
  {Yu}(1995)}]{PhysRevA.51.1845}%
  \BibitemOpen
  \bibfield  {author} {\bibinfo {author} {\bibfnamefont {C.-P.}\ \bibnamefont
  {Sun}}\ and\ \bibinfo {author} {\bibfnamefont {L.-H.}\ \bibnamefont {Yu}},\
  }\href {\doibase 10.1103/PhysRevA.51.1845} {\bibfield  {journal} {\bibinfo
  {journal} {Phys. Rev. A}\ }\textbf {\bibinfo {volume} {51}},\ \bibinfo
  {pages} {1845} (\bibinfo {year} {1995})}\BibitemShut {NoStop}%
\bibitem [{\citenamefont {Albrecht}(1975)}]{Albrecht:1975mtm}%
  \BibitemOpen
  \bibfield  {author} {\bibinfo {author} {\bibfnamefont {K.}~\bibnamefont
  {Albrecht}},\ }\href {\doibase 10.1016/0370-2693(75)90283-X} {\bibfield
  {journal} {\bibinfo  {journal} {Phys. Lett. B}\ }\textbf {\bibinfo {volume}
  {56}},\ \bibinfo {pages} {127} (\bibinfo {year} {1975})}\BibitemShut
  {NoStop}%
\bibitem [{\citenamefont {Gisin}(1982)}]{GISIN1982364}%
  \BibitemOpen
  \bibfield  {author} {\bibinfo {author} {\bibfnamefont {N.}~\bibnamefont
  {Gisin}},\ }\href {\doibase https://doi.org/10.1016/0378-4371(82)90101-7}
  {\bibfield  {journal} {\bibinfo  {journal} {Phys. A}\ }\textbf {\bibinfo
  {volume} {111}},\ \bibinfo {pages} {364} (\bibinfo {year}
  {1982})}\BibitemShut {NoStop}%
\bibitem [{\citenamefont {Kostin}(1972)}]{10.1063/1.1678812}%
  \BibitemOpen
  \bibfield  {author} {\bibinfo {author} {\bibfnamefont {M.~D.}\ \bibnamefont
  {Kostin}},\ }\href {\doibase 10.1063/1.1678812} {\bibfield  {journal}
  {\bibinfo  {journal} {{J. Chem. Phys.}}\ }\textbf {\bibinfo {volume} {57}},\
  \bibinfo {pages} {3589} (\bibinfo {year} {1972})}\BibitemShut {NoStop}%
\bibitem [{\citenamefont {Kadanoff}\ and\ \citenamefont {Baym}(1961)}]{KBBook}%
  \BibitemOpen
  \bibfield  {author} {\bibinfo {author} {\bibfnamefont {L.}~\bibnamefont
  {Kadanoff}}\ and\ \bibinfo {author} {\bibfnamefont {G.}~\bibnamefont
  {Baym}},\ }\href@noop {} {\emph {\bibinfo {title} {{Quantum Statistical
  Mechanics}}}}\ (\bibinfo  {publisher} {The Benjamin/Cummings Publishing
  Company},\ \bibinfo {address} {New York},\ \bibinfo {year}
  {1961})\BibitemShut {NoStop}%
\bibitem [{\citenamefont {Schwinger}(1961)}]{10.1063/1.1703727}%
  \BibitemOpen
  \bibfield  {author} {\bibinfo {author} {\bibfnamefont {J.}~\bibnamefont
  {Schwinger}},\ }\href {\doibase 10.1063/1.1703727} {\bibfield  {journal}
  {\bibinfo  {journal} {J. Math. Phys.}\ }\textbf {\bibinfo {volume} {2}},\
  \bibinfo {pages} {407} (\bibinfo {year} {1961})}\BibitemShut {NoStop}%
\bibitem [{\citenamefont {Rais}\ \emph {et~al.}(2022)\citenamefont {Rais},
  \citenamefont {van Hees},\ and\ \citenamefont {Greiner}}]{Rais:2022}%
  \BibitemOpen
  \bibfield  {author} {\bibinfo {author} {\bibfnamefont {J.}~\bibnamefont
  {Rais}}, \bibinfo {author} {\bibfnamefont {H.}~\bibnamefont {van Hees}}, \
  and\ \bibinfo {author} {\bibfnamefont {C.}~\bibnamefont {Greiner}},\ }\href
  {\doibase 10.1103/PhysRevC.106.064004} {\bibfield  {journal} {\bibinfo
  {journal} {Phys. Rev. C}\ }\textbf {\bibinfo {volume} {106}},\ \bibinfo
  {pages} {064004} (\bibinfo {year} {2022})}\BibitemShut {NoStop}%
\bibitem [{\citenamefont {Keldysh}(1964)}]{Keldysh:1964ud}%
  \BibitemOpen
  \bibfield  {author} {\bibinfo {author} {\bibfnamefont {L.~V.}\ \bibnamefont
  {Keldysh}},\ }\href@noop {} {\bibfield  {journal} {\bibinfo  {journal} {Zh.
  Eksp. Teor. Fiz.}\ }\textbf {\bibinfo {volume} {47}},\ \bibinfo {pages}
  {1515} (\bibinfo {year} {1964})}\BibitemShut {NoStop}%
\bibitem [{\citenamefont
  {Danielewicz}(1984{\natexlab{a}})}]{Danielewicz:1982kk}%
  \BibitemOpen
  \bibfield  {author} {\bibinfo {author} {\bibfnamefont {P.}~\bibnamefont
  {Danielewicz}},\ }\href {\doibase 10.1016/0003-4916(84)90092-7} {\bibfield
  {journal} {\bibinfo  {journal} {Annals Phys.}\ }\textbf {\bibinfo {volume}
  {152}},\ \bibinfo {pages} {239} (\bibinfo {year}
  {1984}{\natexlab{a}})}\BibitemShut {NoStop}%
\bibitem [{\citenamefont {Dahlen}\ \emph {et~al.}(2006)\citenamefont {Dahlen},
  \citenamefont {van Leeuwen},\ and\ \citenamefont
  {Stan}}]{dahlen2006propagating}%
  \BibitemOpen
  \bibfield  {author} {\bibinfo {author} {\bibfnamefont {N.~E.}\ \bibnamefont
  {Dahlen}}, \bibinfo {author} {\bibfnamefont {R.}~\bibnamefont {van Leeuwen}},
  \ and\ \bibinfo {author} {\bibfnamefont {A.}~\bibnamefont {Stan}},\ }\href
  {\doibase 10.1088/1742-6596/35/1/0310} {\bibfield  {journal} {\bibinfo
  {journal} {J. Phys.: Conf. Ser.}\ }\textbf {\bibinfo {volume} {35}},\
  \bibinfo {pages} {340} (\bibinfo {year} {2006})}\BibitemShut {NoStop}%
\bibitem [{\citenamefont {Dahlen}\ and\ \citenamefont {van
  Leeuwen}(2007)}]{dahlen2007solving}%
  \BibitemOpen
  \bibfield  {author} {\bibinfo {author} {\bibfnamefont {N.~E.}\ \bibnamefont
  {Dahlen}}\ and\ \bibinfo {author} {\bibfnamefont {R.}~\bibnamefont {van
  Leeuwen}},\ }\href@noop {} {\  (\bibinfo {year} {2007})},\ \Eprint
  {http://arxiv.org/abs/cond-mat/0703411} {arXiv:cond-mat/0703411
  [cond-mat.mtrl-sci]} \BibitemShut {NoStop}%
\bibitem [{\citenamefont {Stan}\ \emph {et~al.}(2009)\citenamefont {Stan},
  \citenamefont {Dahlen},\ and\ \citenamefont {van Leeuwen}}]{Stan_2009}%
  \BibitemOpen
  \bibfield  {author} {\bibinfo {author} {\bibfnamefont {A.}~\bibnamefont
  {Stan}}, \bibinfo {author} {\bibfnamefont {N.~E.}\ \bibnamefont {Dahlen}}, \
  and\ \bibinfo {author} {\bibfnamefont {R.}~\bibnamefont {van Leeuwen}},\
  }\href {\doibase 10.1063/1.3127247} {\bibfield  {journal} {\bibinfo
  {journal} {J. Chem. Phys.}\ }\textbf {\bibinfo {volume} {130}},\ \bibinfo
  {pages} {224101} (\bibinfo {year} {2009})}\BibitemShut {NoStop}%
\bibitem [{\citenamefont
  {Danielewicz}(1984{\natexlab{b}})}]{DANIELEWICZ1984305}%
  \BibitemOpen
  \bibfield  {author} {\bibinfo {author} {\bibfnamefont {P.}~\bibnamefont
  {Danielewicz}},\ }\href {\doibase
  https://doi.org/10.1016/0003-4916(84)90093-9} {\bibfield  {journal} {\bibinfo
   {journal} {Annals of Physics}\ }\textbf {\bibinfo {volume} {152}},\ \bibinfo
  {pages} {305} (\bibinfo {year} {1984}{\natexlab{b}})}\BibitemShut {NoStop}%
\bibitem [{\citenamefont {Köhler}\ \emph {et~al.}(1999)\citenamefont
  {Köhler}, \citenamefont {Kwong},\ and\ \citenamefont
  {Yousif}}]{KOHLER1999123}%
  \BibitemOpen
  \bibfield  {author} {\bibinfo {author} {\bibfnamefont {H.~S.}\ \bibnamefont
  {Köhler}}, \bibinfo {author} {\bibfnamefont {N.~H.}\ \bibnamefont {Kwong}},
  \ and\ \bibinfo {author} {\bibfnamefont {H.~A.}\ \bibnamefont {Yousif}},\
  }\href {\doibase https://doi.org/10.1016/S0010-4655(99)00260-X} {\bibfield
  {journal} {\bibinfo  {journal} {Comp. Phys. Comm.}\ }\textbf {\bibinfo
  {volume} {123}},\ \bibinfo {pages} {123} (\bibinfo {year}
  {1999})}\BibitemShut {NoStop}%
\bibitem [{\citenamefont {Juchem}\ \emph {et~al.}(2004)\citenamefont {Juchem},
  \citenamefont {Cassing},\ and\ \citenamefont {Greiner}}]{Juchem_2004}%
  \BibitemOpen
  \bibfield  {author} {\bibinfo {author} {\bibfnamefont {S.}~\bibnamefont
  {Juchem}}, \bibinfo {author} {\bibfnamefont {W.}~\bibnamefont {Cassing}}, \
  and\ \bibinfo {author} {\bibfnamefont {C.}~\bibnamefont {Greiner}},\ }\href
  {\doibase 10.1103/physrevd.69.025006} {\bibfield  {journal} {\bibinfo
  {journal} {Physical Review D}\ }\textbf {\bibinfo {volume} {69}} (\bibinfo
  {year} {2004}),\ 10.1103/physrevd.69.025006}\BibitemShut {NoStop}%
\bibitem [{\citenamefont {Puig~von Friesen}\ \emph {et~al.}(2010)\citenamefont
  {Puig~von Friesen}, \citenamefont {Verdozzi},\ and\ \citenamefont
  {Almbladh}}]{PhysRevB.82.155108}%
  \BibitemOpen
  \bibfield  {author} {\bibinfo {author} {\bibfnamefont {M.}~\bibnamefont
  {Puig~von Friesen}}, \bibinfo {author} {\bibfnamefont {C.}~\bibnamefont
  {Verdozzi}}, \ and\ \bibinfo {author} {\bibfnamefont {C.-O.}\ \bibnamefont
  {Almbladh}},\ }\href {\doibase 10.1103/PhysRevB.82.155108} {\bibfield
  {journal} {\bibinfo  {journal} {Phys. Rev. B}\ }\textbf {\bibinfo {volume}
  {82}},\ \bibinfo {pages} {155108} (\bibinfo {year} {2010})}\BibitemShut
  {NoStop}%
\bibitem [{\citenamefont {Schenke}\ and\ \citenamefont
  {Greiner}(2006)}]{Schenke:2005ry}%
  \BibitemOpen
  \bibfield  {author} {\bibinfo {author} {\bibfnamefont {B.}~\bibnamefont
  {Schenke}}\ and\ \bibinfo {author} {\bibfnamefont {C.}~\bibnamefont
  {Greiner}},\ }\href {\doibase 10.1103/PhysRevC.73.034909} {\bibfield
  {journal} {\bibinfo  {journal} {Phys. Rev. C}\ }\textbf {\bibinfo {volume}
  {73}},\ \bibinfo {pages} {034909} (\bibinfo {year} {2006})},\ \Eprint
  {http://arxiv.org/abs/hep-ph/0509026} {arXiv:hep-ph/0509026} \BibitemShut
  {NoStop}%
\bibitem [{\citenamefont {Meirinhos}\ \emph {et~al.}(2022)\citenamefont
  {Meirinhos}, \citenamefont {Kajan}, \citenamefont {Kroha},\ and\
  \citenamefont {Bode}}]{Meirinhos_2022}%
  \BibitemOpen
  \bibfield  {author} {\bibinfo {author} {\bibfnamefont {F.}~\bibnamefont
  {Meirinhos}}, \bibinfo {author} {\bibfnamefont {M.}~\bibnamefont {Kajan}},
  \bibinfo {author} {\bibfnamefont {J.}~\bibnamefont {Kroha}}, \ and\ \bibinfo
  {author} {\bibfnamefont {T.}~\bibnamefont {Bode}},\ }\href {\doibase
  10.21468/scipostphyscore.5.2.030} {\bibfield  {journal} {\bibinfo  {journal}
  {{SciPost} Physics Core}\ }\textbf {\bibinfo {volume} {5}} (\bibinfo {year}
  {2022}),\ 10.21468/scipostphyscore.5.2.030}\BibitemShut {NoStop}%
\bibitem [{\citenamefont {Balzer}(2011)}]{BalzerPHD}%
  \BibitemOpen
  \bibfield  {author} {\bibinfo {author} {\bibfnamefont {K.}~\bibnamefont
  {Balzer}},\ }\emph {\bibinfo {title} {Solving the Two-time Kadanoff-Baym
  Equations. Application to Model Atoms and Molecules}},\ \href@noop {} {Ph.D.
  thesis},\ \bibinfo  {school} {Christian-Albrechts-Universität zu Kiel}
  (\bibinfo {year} {2011})\BibitemShut {NoStop}%
\bibitem [{\citenamefont {Balzer}\ \emph {et~al.}(2010)\citenamefont {Balzer},
  \citenamefont {Bauch},\ and\ \citenamefont {Bonitz}}]{PhysRevA.82.033427}%
  \BibitemOpen
  \bibfield  {author} {\bibinfo {author} {\bibfnamefont {K.}~\bibnamefont
  {Balzer}}, \bibinfo {author} {\bibfnamefont {S.}~\bibnamefont {Bauch}}, \
  and\ \bibinfo {author} {\bibfnamefont {M.}~\bibnamefont {Bonitz}},\ }\href
  {\doibase 10.1103/PhysRevA.82.033427} {\bibfield  {journal} {\bibinfo
  {journal} {Phys. Rev. A}\ }\textbf {\bibinfo {volume} {82}},\ \bibinfo
  {pages} {033427} (\bibinfo {year} {2010})}\BibitemShut {NoStop}%
\bibitem [{\citenamefont {Baym}\ and\ \citenamefont
  {Kadanoff}(1961)}]{PhysRev.124.287}%
  \BibitemOpen
  \bibfield  {author} {\bibinfo {author} {\bibfnamefont {G.}~\bibnamefont
  {Baym}}\ and\ \bibinfo {author} {\bibfnamefont {L.~P.}\ \bibnamefont
  {Kadanoff}},\ }\href {\doibase 10.1103/PhysRev.124.287} {\bibfield  {journal}
  {\bibinfo  {journal} {Phys. Rev.}\ }\textbf {\bibinfo {volume} {124}},\
  \bibinfo {pages} {287} (\bibinfo {year} {1961})}\BibitemShut {NoStop}%
\bibitem [{\citenamefont {Baym}(1962)}]{PhysRev.127.1391}%
  \BibitemOpen
  \bibfield  {author} {\bibinfo {author} {\bibfnamefont {G.}~\bibnamefont
  {Baym}},\ }\href {\doibase 10.1103/PhysRev.127.1391} {\bibfield  {journal}
  {\bibinfo  {journal} {Phys. Rev.}\ }\textbf {\bibinfo {volume} {127}},\
  \bibinfo {pages} {1391} (\bibinfo {year} {1962})}\BibitemShut {NoStop}%
\bibitem [{\citenamefont {Danielewicz}(1990)}]{DANIELEWICZ1990154}%
  \BibitemOpen
  \bibfield  {author} {\bibinfo {author} {\bibfnamefont {P.}~\bibnamefont
  {Danielewicz}},\ }\href {\doibase
  https://doi.org/10.1016/0003-4916(90)90204-2} {\bibfield  {journal} {\bibinfo
   {journal} {Annals of Physics}\ }\textbf {\bibinfo {volume} {197}},\ \bibinfo
  {pages} {154} (\bibinfo {year} {1990})}\BibitemShut {NoStop}%
\bibitem [{\citenamefont {Millington}\ and\ \citenamefont
  {Pilaftsis}(2013)}]{PhysRevD.88.085009}%
  \BibitemOpen
  \bibfield  {author} {\bibinfo {author} {\bibfnamefont {P.}~\bibnamefont
  {Millington}}\ and\ \bibinfo {author} {\bibfnamefont {A.}~\bibnamefont
  {Pilaftsis}},\ }\href {\doibase 10.1103/PhysRevD.88.085009} {\bibfield
  {journal} {\bibinfo  {journal} {Phys. Rev. D}\ }\textbf {\bibinfo {volume}
  {88}},\ \bibinfo {pages} {085009} (\bibinfo {year} {2013})}\BibitemShut
  {NoStop}%
\bibitem [{\citenamefont {Neidig}(2023)}]{workinprogress}%
  \BibitemOpen
  \bibfield  {author} {\bibinfo {author} {\bibfnamefont {T.}~\bibnamefont
  {Neidig}},\ }\emph {\bibinfo {title} {{Work in progress}}},\ \href@noop {}
  {Ph.D. thesis} (\bibinfo {year} {2023})\BibitemShut {NoStop}%
\bibitem [{\citenamefont {Danielewicz}\ and\ \citenamefont
  {Bertsch}(1991)}]{Danielewicz:1991dh}%
  \BibitemOpen
  \bibfield  {author} {\bibinfo {author} {\bibfnamefont {P.}~\bibnamefont
  {Danielewicz}}\ and\ \bibinfo {author} {\bibfnamefont {G.~F.}\ \bibnamefont
  {Bertsch}},\ }\href {\doibase 10.1016/0375-9474(91)90541-D} {\bibfield
  {journal} {\bibinfo  {journal} {Nucl. Phys. A}\ }\textbf {\bibinfo {volume}
  {533}},\ \bibinfo {pages} {712} (\bibinfo {year} {1991})}\BibitemShut
  {NoStop}%
\bibitem [{\citenamefont {Vovchenko}\ \emph {et~al.}(2020)\citenamefont
  {Vovchenko}, \citenamefont {Gallmeister}, \citenamefont {Schaffner-Bielich},\
  and\ \citenamefont {Greiner}}]{Vovchenko:2019aoz}%
  \BibitemOpen
  \bibfield  {author} {\bibinfo {author} {\bibfnamefont {V.}~\bibnamefont
  {Vovchenko}}, \bibinfo {author} {\bibfnamefont {K.}~\bibnamefont
  {Gallmeister}}, \bibinfo {author} {\bibfnamefont {J.}~\bibnamefont
  {Schaffner-Bielich}}, \ and\ \bibinfo {author} {\bibfnamefont
  {C.}~\bibnamefont {Greiner}},\ }\href {\doibase
  10.1016/j.physletb.2019.135131} {\bibfield  {journal} {\bibinfo  {journal}
  {Phys. Lett. B}\ }\textbf {\bibinfo {volume} {800}},\ \bibinfo {pages}
  {135131} (\bibinfo {year} {2020})},\ \Eprint
  {http://arxiv.org/abs/1903.10024} {arXiv:1903.10024 [hep-ph]} \BibitemShut
  {NoStop}%
\bibitem [{\citenamefont {Neidig}\ \emph {et~al.}(2022)\citenamefont {Neidig},
  \citenamefont {Gallmeister}, \citenamefont {Greiner}, \citenamefont
  {Bleicher},\ and\ \citenamefont {Vovchenko}}]{Neidig:2021bal}%
  \BibitemOpen
  \bibfield  {author} {\bibinfo {author} {\bibfnamefont {T.}~\bibnamefont
  {Neidig}}, \bibinfo {author} {\bibfnamefont {K.}~\bibnamefont {Gallmeister}},
  \bibinfo {author} {\bibfnamefont {C.}~\bibnamefont {Greiner}}, \bibinfo
  {author} {\bibfnamefont {M.}~\bibnamefont {Bleicher}}, \ and\ \bibinfo
  {author} {\bibfnamefont {V.}~\bibnamefont {Vovchenko}},\ }\href {\doibase
  10.1016/j.physletb.2022.136891} {\bibfield  {journal} {\bibinfo  {journal}
  {Phys. Lett. B}\ }\textbf {\bibinfo {volume} {827}},\ \bibinfo {pages}
  {136891} (\bibinfo {year} {2022})},\ \Eprint
  {http://arxiv.org/abs/2108.13151} {arXiv:2108.13151 [hep-ph]} \BibitemShut
  {NoStop}%
\bibitem [{\citenamefont {Coci}\ \emph {et~al.}(2023)\citenamefont {Coci},
  \citenamefont {Gl\"a\ss{}el}, \citenamefont {Kireyeu}, \citenamefont
  {Aichelin}, \citenamefont {Blume}, \citenamefont {Bratkovskaya},
  \citenamefont {Kolesnikov},\ and\ \citenamefont {Voronyuk}}]{Coci:2023daq}%
  \BibitemOpen
  \bibfield  {author} {\bibinfo {author} {\bibfnamefont {G.}~\bibnamefont
  {Coci}}, \bibinfo {author} {\bibfnamefont {S.}~\bibnamefont {Gl\"a\ss{}el}},
  \bibinfo {author} {\bibfnamefont {V.}~\bibnamefont {Kireyeu}}, \bibinfo
  {author} {\bibfnamefont {J.}~\bibnamefont {Aichelin}}, \bibinfo {author}
  {\bibfnamefont {C.}~\bibnamefont {Blume}}, \bibinfo {author} {\bibfnamefont
  {E.}~\bibnamefont {Bratkovskaya}}, \bibinfo {author} {\bibfnamefont
  {V.}~\bibnamefont {Kolesnikov}}, \ and\ \bibinfo {author} {\bibfnamefont
  {V.}~\bibnamefont {Voronyuk}},\ }\href {\doibase 10.1103/PhysRevC.108.014902}
  {\bibfield  {journal} {\bibinfo  {journal} {Phys. Rev. C}\ }\textbf {\bibinfo
  {volume} {108}},\ \bibinfo {pages} {014902} (\bibinfo {year} {2023})},\
  \Eprint {http://arxiv.org/abs/2303.02279} {arXiv:2303.02279 [nucl-th]}
  \BibitemShut {NoStop}%
\bibitem [{\citenamefont {Beraudo}\ \emph {et~al.}(2010)\citenamefont
  {Beraudo}, \citenamefont {Blaizot}, \citenamefont {Faccioli},\ and\
  \citenamefont {Garberoglio}}]{Beraudo:2010tw}%
  \BibitemOpen
  \bibfield  {author} {\bibinfo {author} {\bibfnamefont {A.}~\bibnamefont
  {Beraudo}}, \bibinfo {author} {\bibfnamefont {J.~P.}\ \bibnamefont
  {Blaizot}}, \bibinfo {author} {\bibfnamefont {P.}~\bibnamefont {Faccioli}}, \
  and\ \bibinfo {author} {\bibfnamefont {G.}~\bibnamefont {Garberoglio}},\
  }\href {\doibase 10.1016/j.nuclphysa.2010.06.007} {\bibfield  {journal}
  {\bibinfo  {journal} {Nucl. Phys. A}\ }\textbf {\bibinfo {volume} {846}},\
  \bibinfo {pages} {104} (\bibinfo {year} {2010})},\ \Eprint
  {http://arxiv.org/abs/1005.1245} {arXiv:1005.1245 [hep-ph]} \BibitemShut
  {NoStop}%
\end{thebibliography}%

\end{document}